\documentclass[aps,reprint]{revtex4-1}
\usepackage{mathtools} 
\usepackage{amsmath, amssymb} 
\usepackage{amsthm}
\usepackage{enumerate}  % lets you do \begin{enumerate}[(a)] and such
\usepackage{url} % formats url's nicely using \url{}

\newcommand\cx{{\mathbb C}}% complexes

\newcommand\ints{{\mathbb Z}}

\DeclarePairedDelimiter\abs{\lvert}{\rvert}%
\DeclarePairedDelimiter\norm{\lVert}{\rVert}%

\makeatletter
\let\oldabs\abs
\def\abs{\@ifstar{\oldabs}{\oldabs*}}
\let\oldnorm\norm
\def\norm{\@ifstar{\oldnorm}{\oldnorm*}}
\makeatother

\newcommand\pmat[1]{\begin{pmatrix} #1 \end{pmatrix}}

\newcommand{\bra}[1]{{\left\langle{#1}\right\vert}}
\newcommand{\ket}[1]{{\left\vert{#1}\right\rangle}}
\newcommand{\braket}[2]{{\langle{#1}\vert{#2}\rangle}}
\newcommand*\pFqskip{8mu}
\catcode`,\active
\newcommand*\pFq{\begingroup
        \catcode`\,\active
        \def ,{\mskip\pFqskip\relax}%
        \dopFq
}
\catcode`\,12
\def\dopFq#1#2#3#4#5{%
        {}_{#1}F_{#2}\biggl(\genfrac..{0pt}{}{#3}{#4};#5\biggr)%
        \endgroup
}

\begin{document}
\title{Perfect State Transfer in a Spin Chain without Mirror Symmetry}
\author{Gabriel Coutinho$^1$}

\author{Luc Vinet$^2$}
\author{Hanmeng Zhan$^2$}
\author{Alexei Zhedanov$^3$}
\address{$^1$Federal University of Minas Gerais, Belo Horizonte, MG, Brazil}
\address{$^2$Centre de Recherches Math\'ematiques
Universit\'e de Montr\'eal, P.O. Box 6128, Centre-ville Station,
Montr\'eal, Qu\'ebec, Canada, H3C 3J7}
\address{$^3$Department of Mathematics, School of Information, Renmin University of China, Beijing 100872, China}

\begin{abstract}
We introduce an analytical $XX$ spin chain with asymmetrical transport properties. It has an even number $N+1$ of sites labeled by $n=0,\cdots N$. It does not exhibit perfect state transfer (PST) from end-to-end but rather from the first site to the next to last one. In fact, PST of one-excitation states takes place between the even sites: $n\leftrightarrow N-n-1$, $n=0,2,\cdots, N-1$; while states localized at a single odd site undergo fractional revival (FR) over odd sites only. Perfect return is witnessed at double the PST/FR time. The couplings and local magnetic fields are related to the recurrence coefficients of the dual -1 Hahn polynomials.
\end{abstract}

\maketitle

\section{Introduction}
The use of spin chains as devices to implement quantum information tasks is being thoroughly examined. One motivation is that external controls are minimized by calling upon chain dynamics to realize desired circuits. It has been found in particular that spin systems can be engineered to produce perfect state transfer (PST) \cite{Bose2003,Christandl2004,Burgarth2005,Christandl2005,Albanese2004,Yung2005,Karbach2005,Kay2010,Vinet2012a}, that is to transport a single qubit from one site to another with probability one. Another function is the generation of entanglement which results when spin chains yield fractional revival (FR) \cite{Dai2010,Banchi2015,Genest2016,Christandl2017}, that is when they evolve a qubit into a superposition of localized states. The creation of GHZ states can also be accomplished \cite{Clark2005,Kay2007,Kay2018} and the range of one-excitation states that can be obtained dynamically from an excitation initially located at a single site has been looked at \cite{Kay2017,Kay2017a}.

In most studies up to now, it has been assumed that the spin chain is mirror-symmetric, in other words that the Hamiltonian is invariant under reflection with respect to the center of the chain. This is in fact necessary to have PST between the extremities of the chain \cite{Kay2010}. While a five-site example was given in \cite[Section IV.B]{Kay2011} with asymmetric transfer, we here present a system where PST occurs in the absence of this mirror symmetry for an arbitrary (even) number of sites.

Analytic solutions are always attractive in view of their elegance and because they provide users with closed formulas \cite{Bosse2017}. Finding systems with specific transport properties that are amenable to exact treatment is hence quite pertinent. It is with this perspective that we are reporting the discovery of an analytic model that shows PST between its first site and the one that is next to last. For practical purposes, this perfect state transfer could prove as useful as the end-to-end one and might in fact have advantages in certain circumstances when the desired target site is not physically located in the extremity of the chain, and while the sites are too closely located to allow for a complete uncoupling of the last one. In fact, we believe we are contributing to the development of a rich theory of state manipulation in quantum networks, that shall eventually lead to having very good analytic understanding of the transport tasks which are theoretically achievable. 

Each of the spin chains we exhibit in this paper has an even total number $N+1$ of sites and satisfies the following interesting features:
\begin{enumerate}[(i)]
\item it exhibits PST between reciprocal even sites $n$ and $N-n-1$, $n=0,2,\cdots,N-1$;
\item it induces fractional revival of a single qubit initially located on any odd site with the resurgences occurring only on the odd sites;
\item it shows perfect return at double the PST or FR times.
\end{enumerate}

The paper will unfold as follows. We shall start with generalities regarding the one-excitation dynamics of $XX$ spin chains and their connections to orthogonal polynomials. We shall then describe the conditions on the one-excitation spectrum and the associated polynomials for PST to take place between one end of the chain and any other site, subsequently we shall introduce a newly-found model based on the family of dual -1 Hahn polynomials and show that it realizes the conditions for PST between the first and the next to last sites. After observations on regularities in the set of coupling constants and local magnetic fields, we shall proceed to demonstrate that this novel analytic spin chain possesses the remarkable transport properties that we mentioned above.

The couplings and magnetic fields of our chains are given in (\ref{eqns_strength}), with integer $\xi=\eta+1$ and PST time $\pi/4$.

\section{Generalities}

We shall be concerned with a spin chain of the $XX$-type with a Hamiltonian $H$ of the form 

\begin{equation}
H = \frac{1}{2}\sum_{\ell=0}^{N-1} J_{\ell+1} (\sigma_{\ell}^x\sigma_{\ell+1}^x + \sigma_{\ell}^y\sigma_{\ell+1}^y) + \frac{1}{2} \sum_{\ell=0}^N B_{\ell} (\sigma_{\ell}^z+1)
\end{equation}
with $J_{\ell}>0$ the coupling constants and $B_{\ell}$ the local magnetic fields. The boundary conditions $J_0=J_{N+1}=0$ are assumed. This operator acts on $(\cx^2)^{\otimes (N+1)}$. The symbols $\sigma_{\ell}^x$, $\sigma_{\ell}^y$, $\sigma_{\ell}^z$ stand as usual for the Pauli matrices with the index $\ell$ indicating on which of the $(N+1)$ $\cx^2$ factors they act. We shall denote by $\ket{\uparrow}$ and $\ket{\downarrow}$ the eigenvectors of $\sigma^z$ with eigenvalue $1$ and $-1$, respectively. We shall focus on the $1$-excitation states $\ket{\downarrow}^{\otimes (n-1)}\ket{\uparrow}\ket{\downarrow}^{\otimes (N-n+1)}$, $n=0,\cdots, N$, which can be identified with the unit vectors
\begin{equation}
\ket{n} = (0,\cdots,1,\cdots, 0,0)
\label{eqn_ketn}
\end{equation}
in $\cx^{N+1}$ with the single $1$ in the $n$-th entry. It is readily seen that $H$ leaves their linear span invariant and is given in that basis by the Jacobi matrix
\begin{equation}
J = \pmat{B_0 & J_1 &  &  & \\
J_1 & B_1 & J_2 &  & \\
 & \ddots & \ddots & \ddots & \\
 &  & J_{N-1} & B_{N-1} & J_N\\
 &  &  & J_N & B_N} .
\label{eqn_Jac}
\end{equation}
Such matrices are diagonalized by orthogonal polynomials. Expanding the normalized eigenstates $\ket{x_s}$ of $J$ with eigenvalues $x_s$, $s=0,\cdots,N$ over the occupation basis:
\begin{equation}
\ket{x_s} = \sum_{n=0}^N \sqrt{w_s} \chi_n(x_s)\ket{n},
\end{equation}
where $w_s$ are the appropriate normalizing constants, we shall have $J\ket{x_s} = x_s \ket{x_s}$ provided $\chi_n(x)$ are orthogonal polynomials \cite{Chihara2011} satisfying the three-term recurrence relation:
\begin{equation}
x \chi_n(x) = J_{n+1}\chi_{n+1}(x)+B_n\chi_n(x)+J_n\chi_{n-1}(x).
\label{eqn_xschin}
\end{equation}
As the eigenstates are normalized, we also have
\begin{equation}
\ket{n} = \sum_{s=0}^N \sqrt{w_s} \chi_n(x_s) \ket{x_s},
\label{eqn_expand}
\end{equation}
which implies that $w_s$ plays the role of the weight function:
\begin{equation}
\sum_{s=0}^N w_s\chi_n(x_s)\chi_m(x_s) = \delta_{nm}.
\end{equation}
(The coefficients $\chi_n(x_s)$ are taken to be real.) We shall also make use of the monic version $P_n(x)$ of the polynomials $\chi_n(x)$ which are normalized such that $P_n(x)=x^n+\cdots$, i.e. such that the coefficient of the leading monomial is $1$. The relation between $P_n(x)$ and $\chi_n(x)$ is
\begin{equation}
P_n(x) = \sqrt{h_n}\chi_n(x)
\label{eqn_Pn}
\end{equation}
with $h_n=J_1^2J_2^2\cdots J_n^2$, and we have
\begin{equation}
xP_n(x) = P_{n+1}(x) + b_n P_n(x) + u_n P_{n-1}(x)
\label{eqn_rec}
\end{equation}
with $b_n=B_n$ and $u_n=J_n^2$ which confirms the monic property.

Because of the conditions $J_i>0$, the eigenvalues $x_s$ of $J$ are distinct; we shall assume in the following that they are ordered $x_0<x_1<\cdots<x_N$. The characteristic polynomial is
\begin{equation}
P_{N+1}(x)=(x-x_0)\cdots(x-x_N).
\end{equation}
It is known \cite{Chihara2011} that the weights $w_s$ are given by
\begin{equation}
w_s = \frac{h_N}{P_N(x_s)P_{N+1}'(x_s)}
\label{eqn_ws}
\end{equation}
where $P_{N+1}'(x) = \frac{d}{dx} P_{N+1}(x)$. 

\section{Conditions for PST}

At this point, for simplicity, we shall suppose that the single qubit is initially located at the site $0$. To have PST of this qubit at time $t=T$ to the site $n$ requires that
\begin{equation}
e^{-iTH} \ket{0} = e^{i\phi} \ket{n}
\label{eqn_pst}
\end{equation}
where $\phi$ is an arbitraty phase. Since $\chi_0(x)=1$, given (\ref{eqn_ketn}), this implies the spectral condition
\begin{equation}
e^{-iTx_s} = e^{i\phi} \chi_n(x_s).
\label{eqn_pstev}
\end{equation}
A necessary condition for this PST readily follows, namely
\begin{equation}
\chi_n(x_s)=\pm 1
\label{eqn_pstchi}
\end{equation}
since the polynomials $\chi_n(x_s)$ are real. If we define $\sigma_n(s)\in \ints$ so that $\chi_n(x_s) = (-1)^{\sigma_n(s)}$, we may translate the relation (\ref{eqn_pst}) into
\begin{equation}
T x_s = -\phi + \pi\sigma_n(s) + 2\pi M_n(s),
\label{eqn_pstMn}
\end{equation}
where $M_n(s)$ is a series of integers that may depend on $s$ (and on $n$ as well).

Finding a chain with such a PST property therefore amounts to finding a family of polynomials $\chi_n(x)$ orthogonal on a discrete grid $x_s$ and such that first (\ref{eqn_pstchi}) and then (\ref{eqn_pstMn}) are satisfied. The recurrence coefficients of these polynomials will hence provide the couplings and magnetic fields and will thereby determine the Hamiltonian.

In the familiar case of PST from end-to-end, we have $n=N$ and (\ref{eqn_pstchi}) involves the polynomial $\chi_N$ of degree $N$. Since the zeros of $\chi_N(x)$ interlace those of $P_{N+1}(X)$ at $x=x_s$, we conclude that the $1$ and $-1$ of (\ref{eqn_pstchi}) must alternate between consecutive grid points. Moreover, since the sign of $P_{N+1}'(x_s)$ goes like $(-1)^{N+s}$ when the eigenvalues are ordered, in view of (\ref{eqn_ws}), we must have $\sigma_N(x)=N+s$ for the weights $w_s$ to be positive. We thus retrieve known results for PST between the extremities. The condition $\chi_N(x_s)=(-1)^{N+s}$ is in fact tantamount \cite{Albanese2004} to demanding mirror symmetry:
\begin{equation}
J_n = J_{N-n+1},\quad B_n = B_{N-n}.
\end{equation}
Spin chains with end-to-end PST can thus be obtained by looking for (special cases of) orthogonal polynomials with mirror symmetric recurrence coefficients and checking that their orthogonality grid points satisfy \eqref{eqn_pstMn}. A number of analytic models with PST between their extremities have been found in that way; among them \cite{Christandl2004,Albanese2004,Christandl2005,Kay2010,Vinet2012a} is the chain associated to the Krawtchouk polynomials. There is another one initially found in \cite{Shi2005} (see also \cite{Stoilova2011}) which is obtained from the much less known dual -1 Hahn polynomials \cite{Vinet2012,Chihara2011}. We shall be exploiting these polynomials again to find a chain that is not mirror-symmetric and that exhibits for one thing PST between sites $0$ and $N-1$. This will require showing that (\ref{eqn_pstchi}) and (\ref{eqn_pstMn}) with $n=N-1$ are satisfied for this model.

\section{A chain with asymmetric PST}
We shall focus on $XX$ spin chain with an even number of sites labeled by $n=0,\cdots, N$ with $N$ odd and with couplings $J_n$ and magnetic fields $B_n$ given by
\begin{subequations}
\begin{eqnarray}
J_n =2\sqrt{[n]_{\xi} [N-n+1]_{\eta}}\\
B_n =  (-1)^{n+1} 2(\xi - \eta)
\end{eqnarray}
\label{eqns_strength}%
\end{subequations}
where $[m]_{\mu} = m + (1-(-1)^m)\mu$, and $\xi$ and $\eta$ are two real parameters satisfying $\xi, \eta>-1/2$. The corresponding Jacobi matrix (\ref{eqn_Jac}) is diagonalized by the dual -1 Hahn polynomials introduced in \cite{Tsujimoto2013} (see also \cite{Genest2013}). Useful properties of these functions are collected in the Appendix. Essential for our argument is the fact that they are orthogonal on the grid
\begin{equation}
y_s = (-1)^s (2s + 2\xi + 2\eta + 1)+1,\quad s=0,\cdots, N.
\end{equation}
A special case of these chains is already known \cite{Shi2005,Stoilova2011,Vinet2012} to exhibit end-to-end PST. Indeed, if $\xi = \eta$, the chain is readily seen to be mirror-symmetric which implies as already noted that $\chi_N(x_s) = (-1)^{N+s}$. The ordered spectral set $\{x_s\}$ (equivalent as a set to $\{y_s\}$) is given by
\begin{equation}
x_s = \begin{cases}
-4\xi +4s -2N ,\quad s=0,\cdots,\frac{N-1}{2}\\
4\xi + 4s - 2N ,\quad s = \frac{N+1}{2},\cdots,N
\end{cases}
\end{equation}
and consists of two linear sublattices with step $4$ separated by a gap of $8\xi+4$. It is straight forward to see that condition (\ref{eqn_pstMn}) is satisfied for $x_s$ given by (\ref{eqn_xs}) with
\begin{equation}
T = (2k+1) \frac{\pi}{4},\quad k=0,1,2,\cdots
\end{equation}
and $\xi$ integer. This reconfirms that the $XX$ chain with specification (\ref{eqns_strength}) where $\xi=\eta =$ an integer has PST at time $T = \frac{\pi}{4}$. (Note that $B_n=0$ in this case.)

We shall now consider the new situation where
\begin{equation}
\xi = \eta +1.
\end{equation}
Clearly, the corresponding chain is no longer mirror-symmetric. We shall check that it however induces PST between the sites $0$ and $N-1$. To that end, we first need to show that $\chi_{N-1}(x_s) = \pm 1$ and to obtain the corresponding sequence $(-1)^{\sigma_{N-1}(x_s)}$ of $1$s and $-1$s. Here $y_s = (-1)^s(2s + 4\eta + 3)+1$ and the ordered eigenvalues of the Jacobi matrix (\ref{eqn_Jac}) are  %CHECK exponent in y_s (I changed 2 to s)
\begin{equation}
x_s = \begin{cases}
-4\eta + 4s -2N -2,\quad s=0,\cdots,\frac{N-1}{2}\\
4\eta + 4s -2N +2,\quad s=\frac{N+1}{2},\cdots,N.
\label{eqn_xs}
\end{cases}
\end{equation}
With $\delta= \eta+1$, the polynomial $\chi_{N-1}(x)$ is given by
\begin{align}
\begin{split}
\chi_{N-1}(x) &= (-1)^{\frac{N-1}{2}} \pFq{3}{2}{\frac{1-N}{2}, \delta + \frac{x}{4}, \delta - \frac{x}{4}} {\frac{1-N}{2}, \frac{2\eta+3}{2}}{1} \\
&= (-1)^{\frac{N-1}{2}} \pFq{2}{1}{\delta + \frac{x}{4}, \delta-\frac{x}{4}}{\frac{2\eta+3}{2}}{1}
\end{split}
\label{eqns_chiN-1}
\end{align}
owing to the definition of hypergeometric series (see (\ref{eqn_hypergeometric})). The Gauss series in the second equality should be taken to be terminating after $\frac{N-1}{2}$ terms since it is simply obtained from cancelling factors in the terminating $_3F_2$. Definition (\ref{eqns_chiN-1}) can be found using the formulas on dual -1 Hahn polynomials gathered in the Appendix. The factor $(-1)^{\frac{N-1}{2}}$ must be present for the leading term of $\chi_{N-1}(x)$ to be positive.

It follows from (\ref{eqn_xs}) that
\begin{equation}
\chi_{N-1}(x_s) = \begin{cases}
(-1)^s,\quad s = 0,\cdots, \frac{N-1}{2}\\
(-1)^{s+1},\quad s= \frac{N+1}{2},\cdots,N.
%CHECK exponent here (I changed - to +)
\end{cases}\label{eqns_chiN-1cases}
\end{equation}
For example, if $s=0,\cdots, \frac{N-1}{2}$,
\begin{align}
\chi_{N-1}(x_s) = (-1)^{\frac{N-1}{2}}\pFq{2}{1}{s-\frac{N-1}{2}, \frac{N-1}{2}-s+2\eta+2}{\eta+\frac{3}{2}}{1}
\end{align}
and one arrives at the first line of (\ref{eqns_chiN-1cases}) with the help of the Vandermonde summation formula
\begin{align}
\pFq{2}{1}{-n, b}{c}{1} = \frac{(c-b)_n}{(c)_n}
\end{align}
and the simple fact that $(-k-a)_k = (-1)^k(a+1)_k$. The second line of (\ref{eqns_chiN-1cases}) is found in the same way. Having observed that the necessary condition (\ref{eqn_pstchi}) is satisfied with the sequence $\sigma_{N-1}(s)$ given by the exponents of $-1$ in (\ref{eqns_chiN-1cases}), we now turn to (\ref{eqn_pstMn}) which yields
\begin{widetext}
\begin{subequations}
\begin{eqnarray}
T (-4\eta +4s -2N - 2) = -\phi + \pi s + 2\pi (cs+d),\quad s=0,\cdots, \frac{N-1}{2} \label{eqn_T1}\\
T(4\eta +4s -2N +2) = -\phi + \pi\left(s+1\right) +2\pi (cs+e),\quad s = \frac{N+1}{2},\cdots, N \label{eqn_T2}
\end{eqnarray}
\end{subequations}
\end{widetext}
with $c$, $d$, $e$ integers. Equating the coefficients of $s$ and the constant terms in (\ref{eqn_T1}) and in (\ref{eqn_T2}) shows that PST occurs between the sites $0$ and $N-1$ at the times $T=\frac{\pi}{4}(2c+1)$, with $c$ integer, provided $\eta$ is itself an integer. (The phase $\phi$ is fixed by the constant part of either equation once their compatibility has been ensured.)

\section{Remarkable state transport properties}
We shall now confirm that the special PST, FR and return properties we mentioned before are indeed realized in the spin chain without mirror-symmetry we have designed. Recall that this model is specified by the couplings and fields given in (\ref{eqns_strength}) with $\xi = \eta + 1$ with $\eta$ integer.

Let us first underscore some regularities among the corresponding recurrence coefficients $b_n = B_n$ and $u_n = J_n^2$ of the monic polynomials $P_n(x)$.

First we note that 
\begin{equation}
b_n=(-1)^{n+1} 2,\quad n=0,\cdots, N.
\end{equation}
Second, it is directly checked that 
\begin{equation}
u_{2n-1}u_{2n} = u_{N-2n+1}u_{N-2n},\quad n=1,2,\cdots,(N-1)/2
\label{eqn_reccoeff}
\end{equation}
since $[k]_{\eta+1} = [k+2]_{\eta}$ when $k$ is odd.

Consider now the amplitude $A_{\ell m}(T)$ to find at site $\ell$, a qubit initially at site $m$, after a time $T$ equal to the PST time between the sites $0$ and $N-1$:
\begin{align}
\begin{split}
A_{\ell m}(T) &= \bra{\ell}e^{-iTH} \ket{m}\\
&= \sum_{s=0}^N w_s e^{-iT x_s} \chi_{\ell}(x_s) \chi_m(x_s).
\end{split}
\end{align}
The second expression is obtained using the expansion (\ref{eqn_expand}) and $\braket{x_s}{x_{s'}} =\delta_{s s'}$. Note that $A_{\ell m}(T)=A_{m\ell}(T)$. For the model at hand, we know that $e^{-ix_s T}= e^{i\phi} \chi_{N-1}(x_s)$ since we have PST between $0$ and $N-1$. Let us see what this entails for $A_{\ell m}(T)$. The three-term recurrence relation (\ref{eqn_rec}) gives
\begin{equation}
P_{N-1}(x_s) = \frac{x_s-b_N}{u_N} P_N(x_s)
\end{equation}
since $P_{N+1}(x_s)=0$. Using (\ref{eqn_Pn}), the definition of $h_n$ and (\ref{eqn_ws}) we obtain
\begin{equation}
w_s e^{-ix_s T} = e^{i\phi} w_s \chi_{N-1}(x_s) = e^{i\phi}\frac{\sqrt{h_{N-1}}(x_s-b_N)}{{P_{N+1}'}(x_s)}.
\end{equation}
Recalling that the $N$-th order divided difference operators $\Delta^N$ can be defined by 
\begin{equation}
\Delta^N(f(x))=\sum_{s=0}^N \frac{f(x_s)}{P_{N+1}'(x_s)},
\end{equation}
we thus find the following formula:
\begin{equation}
A_{\ell m}(T) = e^{i\phi} \sqrt{h_{N-1}} \Delta^N[(x-b_N) \chi_{\ell}(x) \chi_m(x)].
\label{eqn_Alm}
\end{equation}
It is known that $\Delta^N\phi(x)=0$ for any polynomial of degree smaller than $N$ and that $\Delta^N(x^N)=1$. It is therefore manifest that
\begin{equation}
A_{\ell m}(T)=0,\quad \text{ if } \ell+m < N-1.
\end{equation}
\begin{enumerate}[(i)]
\item PST between even sites

Consider the amplitude $A_{2n, N-2n-1}(T)$ for transfer between the even sites $2n$ and $N-2n-1$. In this case the polynomial $(x-b_N)\chi_{2n}(x) \chi_{N-2n-1}(x)$ arising in (\ref{eqn_Alm}) has degree $N$ with the coefficient of $x^N$ given by $1/\sqrt{h_{2n}h_{N-2n-1}}$. It readily follows that 
\begin{equation}
A_{2n,N-2n-1}(T)= e^{i\phi}\sqrt{\frac{h_{N-1}}{h_{2n}h_{N-2n-1}}} = e^{i\phi}
\end{equation}
upon using the property (\ref{eqn_reccoeff}) of the recurrence coefficients for our special model. Since the modulus of $A_{2n,N-2n-1}(T)$ is one, we thus find that over the time $T$, PST can occur not only between $0$ and $N-1$ but between the even sites $2N$ and $N-2n-1$.

\item FR on odd sites

It follows from unitarity that a qubit initially at an odd site is revived at only odd sites after time $T$.  We provide detailed calculations on the amplitudes.

Let us now examine the state at time $T$ of a qubit initially located at site $1$; that is, let us look at the amplitude $A_{m1}(T) = e^{i\phi} \sqrt{h_{N-1}} \Delta^N ((x-b_N)\chi_1(x) \chi_m(x))$. From (\ref{eqn_xschin}) we have
\begin{equation}
\chi_1(x) = \frac{1}{\sqrt{h_1}} (x-b_0)
\end{equation}
and since $b_0=-2$ and $b_N=2$,
\begin{equation}
(x-b_N)\chi_1(x) = \frac{1}{\sqrt{h_1}} (x^2-4).
\end{equation}
In view of (\ref{eqn_Alm}), $A_{m1}(T)=0$ for $m\ne N-2, N-1, N$. We can rule out a transition from $1$ to the even site $N-1$ for the following reason. It is manifest from the explicit formula (\ref{eqn_PnxxietaN}) for $P_n(x;\eta+1, \eta, N)$ with $n$ even where only factors of the form $(\delta+\frac{x}{4})_k (\delta-\frac{x}{4})_k$ occur, that the polynomial $P_{N-1}$ will only involve monomial of even degrees. Multiplying this polynomial by $(x^2-4)$ will not generate a term of odd degree $N$ and hence $A_{1,N-1}(T)=0$. It follows that after time $T$ there is FR over the sites $N-2$ and $N$ of the qubit initially at site $1$. The amplitude $A_{N-2,1}(T)$ is readily computed to be
\begin{equation}
A_{N-2,1}(T) = e^{i\phi} \sqrt{\frac{h_{N-1}}{h_1 h_{N-2}}} = e^{i\phi} \sqrt{\frac{u_{N-1}}{u_1}}
\end{equation}
so that the probabilities for the transitions $\ket{1} \to \ket{N-2}$ and $\ket{1} \to \ket{N}$ are respectively
\begin{equation}
P_{1\to N-2} = \frac{u_{N-1}}{u_1} \text{ and }  P_{1\to N} = 1- \frac{u_{N-1}}{u_1}.
\end{equation}
One can argue in a similar fashion that a qubit initially at the site $2n+1$ will undergo fractional revival over the odd sites $N,N-2,\cdots,N-2n$ after time $T$.

\item Perfect return

Simple reasoning finally shows that perfect return will be observed at double the PST/FR time $T$. Indeed we know that in the eigenbasis of the Jacobi matrix, 
\begin{equation}
e^{-iTJ} = \mathrm{Diag}(e^{-iTs}) = \mathrm{Diag}(\pm 1),
\end{equation}
It immediately follows that
\begin{equation}
e^{-i2TJ} = [\mathrm{Diag}(e^{-iTs})]^2 =  I
\end{equation}
and hence single qubits come back to their original positions after $2T$.
\end{enumerate}

\section{Conclusion}
We have presented an analytic $XX$ spin chain without mirror symmetry that exhibits interesting transport properties. This chain with an even number of sites exhibits PST between its even sites and FR over its odd sites.

The prospect for experimental realizations seems good and that could probably be achieved in photonic lattice following approaches already employed \cite{PerezLeija2013,Chapman2016}. One might think that a transfer to the next to last site could have merits comparable to end-to-end PST. In any event the results presented here illustrate the wealth of tasks that spin chains can accomplish and suggest further studies.

It is natural to ask if there are other similar examples of analytic spin chains without mirror symmetry and with interesting transport properties. In this respect, one might wish to obtain a characterization of such models and their symmetries and to develop constructive algorithms that could solve the corresponding inverse spectral problems. We plan on following up on these questions.

\begin{acknowledgements}
We have benefited from stimulating discussions with Christino Tamon. Reference \cite{Kay2011} was put to our attention after circulation of the first version of this paper. Two of us (Gabriel Coutinho) and (Alexei Zhedanov) would like to thank the Centre de Recherches Math\'ematiques (CRM) for its hospitality while this work was initiated. The research of Luc Vinet is supported in part by a discovery grant from NSERC (Canada) and that of Alexei Zhedanov by the National Science Foundation of China (Grant Number 11771015).
\end{acknowledgements}

\appendix

\section{the dual -1 Hahn polynomials}
We collect here relevant information on the dual -1 Hahn polynomials \cite{Tsujimoto2013,Genest2013} which in monic form are denoted $P_n(x;\xi,\eta, N)$. They depend on two real parameters $\xi,\eta$ with $\xi,\eta>-1$ and on an integer $N$ and have been introduced as a $q=-1$ limit of the dual $q$-Hahn polynomials \cite{Koekoek2010}. They satisfy the three-term recurrence relation
%\begin{widetext}
\begin{multline}
xP_n(x) =P_{n+1}(x) + (-1)^{n+1} (2\xi + (-1)^N 2\eta) P_n(x) \\
+ 4[n]_{\xi} [N-n+1]_{\eta} P_{n-1}(x)
\end{multline}
%\end{widetext}
with $[n]_{\mu}=n+(1-(-1)^n)\mu$. Note that the variable $x$ has been shifted by $1$ with respect to the definitions in references \cite{Tsujimoto2013,Genest2013}. We shall only record formulas that pertain to the case $N$ odd. Recall that the hypergeometric series $_rF_s$ is defined by
\begin{align}
\pFq{r}{s}{a_1,\cdots,a_r}{b_1,\cdots,b_s}{z} = \sum_{k=0}^{\infty} \frac{(a_1)_k\cdots(a_r)_k}{(b_1)_k\cdots(b_s)_k} {\frac{z^k}{k!}}
\label{eqn_hypergeometric}
\end{align}
with $(c)_k = c(c+1)\cdots(c+k-1)$. For $N$ odd, the dual -1 Hahn polynomials are given by
\begin{widetext}
\begin{align*}
P_n(x; \xi,\eta, N) = \begin{cases}
16^{\frac{n}{2}} \left(\frac{1-N}{2}\right)_{\frac{n}{2}} \left(\frac{2\xi+1}{2}\right)_{\frac{n}{2}} \pFq{3}{2}{-\frac{n}{2}, \delta + \frac{x}{4}, \delta - \frac{x}{4}}{\frac{1-N}{2}, \frac{2\xi+1}{2}}{1},\quad n \text{ even}\\
16^{\frac{n-1}{2}}\left(\frac{1-N}{2}\right)_{\frac{n-1}{2}} \left(\frac{2\xi+3}{2}\right)_{\frac{n-1}{2}} (x+2\xi - 2\eta) \pFq{3}{2}{-\frac{n-1}{2}, \delta+ \frac{x}{4}, \delta -\frac{x}{4}}{\frac{1-N}{2}, \frac{2\xi+3}{2}}{1}, \quad n \text{ odd} \end{cases}  
\label{eqn_PnxxietaN}
\end{align*}
\end{widetext}
where $\delta = \frac{\xi+\eta+1}{2}$.

These polynomials obey an orthogonality relation of the form
\begin{multline}
\sum_{s=0}^N w_s(\xi,\eta, N) P_n(y_s; \xi,\eta, N) P_m(y_s; \xi, \eta, N) \\
= \nu_n (\xi,\eta, N) \delta_{n,m}
\end{multline}
on the grid points
\begin{equation}
y_s = (-1)^s (2s + 2\eta + 2\xi +1) +1.
\end{equation}
The weights and normalization factors are given by
\begin{widetext}
\begin{equation}
\begin{gathered}
w_s(\eta,\xi,N) = \begin{cases}
(-1)^{\frac{s}{2}} \frac{k \left(\frac{1-N}{2}\right)_{\frac{s}{2}} \left(\xi + \frac{1}{2}\right)_{\frac{s}{2}} (\eta + \xi + 1)_{\frac{s}{2}}} {\left(\frac{s}{2}\right)! \left(\eta+\frac{1}{2}\right)_{\frac{s}{2}} \left(\frac{N+3}{2} + \eta + \xi\right)_{\frac{s}{2}}}, \quad s \text{ even}\\
(-1)^{\frac{s}{2}} \frac{k \left(\frac{1-N}{2}\right)_{\frac{s-1}{2}} \left(\xi + \frac{1}{2}\right)_{\frac{s+1}{2}} (\eta + \xi + 1)_{\frac{s-1}{2}}} {\left(\frac{s-1}{2}\right)! \left(\eta+\frac{1}{2}\right)_{\frac{s+1}{2}} \left(\frac{N+3}{2} + \eta + \xi\right)_{\frac{s-1}{2}}}, \quad s \text{ odd}
\end{cases}, \quad \text{ where } k = \frac{\left(\eta+\frac{1}{2}\right)_{\frac{N+1}{2}}}{(\eta + \xi +1)_{\frac{w+1}{2}}},\\
\nu_n(\eta, \xi, N) = \begin{cases}
16^n \left(\frac{n}{2}\right)! \left(\frac{1-N}{2}\right)_{\frac{n}{2}} \left(\xi + \frac{1}{2}\right)_{\frac{n}{2}} \left(\frac{-N}{2} - \eta\right)_{\frac{n}{2}},\quad n \text{ even}\\
-16^n \left(\frac{n-1}{2}\right)! \left(\frac{1-N}{2}\right)_{\frac{n-1}{2}} \left(\xi + \frac{1}{2}\right)_{\frac{n+1}{2}} \left(\frac{-N}{2} - \eta\right)_{\frac{n+1}{2}},\quad n \text{ even}.
\end{cases}
\end{gathered}
\end{equation}
\end{widetext}

\bibliographystyle{apsrev4-1}

\end{document}